\documentclass[letterpaper,cleveref]{lipics-v2019}

\usepackage[round]{natbib}
\usepackage{booktabs}
\usepackage{amsmath,amsthm}
\usepackage{hyperref}
\usepackage{xcolor}
\usepackage{longtable}
\usepackage{paralist}
\usepackage{totcount}
\usepackage{booktabs}

\hypersetup{
colorlinks=true,     
linkcolor=blue,      
citecolor=blue,      
filecolor=magenta,   
urlcolor=cyan        
}

\newif\ifcomments\commentstrue

\ifcomments
\newcommand{\authornote}[3]{\textcolor{#1}{[#3 ---#2]}}
\newcommand{\todo}[1]{\textcolor{red}{[TODO: #1]}}
\else
\newcommand{\authornote}[3]{}
\newcommand{\todo}[1]{}
\fi

\newcounter{totHours} 
\regtotcounter{totHours}

\title{Methodology for Assessing the State of the Practice for Domain X} 
\author{Spencer Smith}{McMaster University, Canada}{smiths@mcmaster.ca}{}{}
\author{Jacques Carette}{McMaster University, Canada}{carette@mcmaster.ca}{}{}
\author{Peter Michalski}{McMaster University, Canada}{michap@mcmaster.ca}{}{}
\author{Ao Dong}{McMaster University, Canada}{donga9@mcmaster.ca}{}{}
\author{Olu Owojaiye}{McMaster University, Canada}{owojaiyo@mcmaster.ca}{}{}

\authorrunning{Smith et al.}  \Copyright{Spencer Smith and Jacques Carette and
 Peter Michalski and Ao Dong and Olu Owojaiye}

\ccsdesc{Software and its engineering}
\ccsdesc{Software and its engineering~Software product lines}
\ccsdesc{General and reference~Empirical studies}

\keywords{software quality, domain analysis, scientific computing software,
  research software, empirical measures, analytic hierarchy process}

\date{\today}

\hideLIPIcs
\nolinenumbers

\begin{document}

\maketitle

\begin{abstract}
	To improve software development methods and tools for research software, we
	first need to understand the current state of the practice.  Therefore, we
	have developed a methodology for assessing the state of the software
	development practices for a given research software domain.  The methodology
	is applied to one domain at a time in recognition that software development in
	different domains is likely to have adopted different best practices.
	Moreover, providing a means to measure different domains facilitates
	comparison of development practices between domains.  For each domain we wish
	to answer questions such as: 
  \begin{inparaenum}[i)]
    \item What artifacts (documents, code, test cases, etc.) are present?
    \item What tools are used?
    \item What principles, process and methodologies are used?
    \item What are the pain points for developers?
    \item What actions are used to improve qualities like maintainability and
    reproducibility?
  \end{inparaenum} 
  To answer these questions, our methodology prescribes the following steps: 
  \begin{inparaenum}[i)] 
    \item Identify the domain;
    \item Identify a list of candidate software packages;
    \item Filter the list to a length of about 30 packages;
    \item Gather source code and documentation for each package;
    \item Collect repository related data on each software package, like number
    of stars, number of open issues, number of lines of code;
    \item Fill in the measurement template (the template consists of 108
    questions to assess 9 qualities (including the qualities of installability,
    usability and visibility));
    \item Interview developers (the interview consists of 20 questions and takes
    about an hour);
    \item Rank the software using the Analytic Hierarchy Process (AHP); and,
    \item Analyze the data to answer the questions posed above.
  \end{inparaenum}
  A domain expert should be engaged throughout the process, to ensure that
  implicit information about the domain is properly represented and to assist
  with conducting an analysis of the commonalities and variabilities between the
  30 selected packages.  Using our methodology, spreadsheet templates and AHP
  tool, we estimate (based on our experience with using the process) the time to
  complete an assessment for a given domain at \total{totHours} person hours.
\end{abstract}

~\newpage

\tableofcontents

~\newpage

\section{Introduction} \label{SecIntroduction}

Research software uses computing to simulate mathematical models of real world
systems so that we can better understand and predict those systems' behaviour.  A
small set of examples of important research software includes the
following: designing new automotive parts, analyzing the flow of blood in the
body, and determining the concentration of a pollutant released into the
groundwater.  As these examples illustrate, research software can be used for
tackling important problems that impact such areas as manufacturing, financial
planning, environmental policy, and the health, welfare and safety of
communities.

Given the importance of research software, scientists and engineers are pushing
for methods and tools to sustainably develop high quality software.  This is
evident from the existence of such groups as the Software Sustainability
Institute (\href{https://www.software.ac.uk/} {SEI}) and Better Scientific
Software (\href{https://bssw.io/} {BSS}).  Sustainability promoting groups such
as these are necessary because unfortunately the current ``state of the
practice'' for research software often does not incorporate ``state of the art''
Software Engineering (SE) tools and methods \citep{JohansonAndHasselbring2018}.
The lack of SE tools and methods contributes to sustainability and reliability
problems \citep{FaulkEtAl2009}.  Problems with the current state of the practice
are evident from embarrassing failures, like a retraction of derived molecular
protein structures \citep{Miller2006}, false reproduction of sonoluminescent
fusion \citep{PostAndVotta2005}, and fixing and then reintroducing the same
error in a large code base three times in 20 year
\citep{MilewiczAndRaybourn2018}.  To improve this situation, we need to first
fully understand the current state of the practice for research software
development.

The purpose of our proposed methodology is to understand how software quality is
impacted by the software development principles, processes and tools currently
used within research software communities.  Since research software is so broad
a category, we will reduce the scope of our methodology to focus on one specific
research software domain at a time.  To emphasize that the method is generic, we
will label the specific domain as X within this document.  

This ``state of the practice for domain X'' exercise builds off of prior work on
measuring/assessing the state of software development practice in several
research domains.  We have updated the work that was done previously for domains
such as Geographic Information Systems \citep{SmithEtAl2018_arXivGIS}, Mesh
Generators \citep{SmithEtAl2016}, Seismology software \citep{SmithEtAl2018}, and
Statistical software for psychology \citep{SmithEtAl2018_StatSoft}.  Initial
tests of the new methodology have been done for medical image analysis software
\citep{Dong2021} and for Lattice Boltzmann Method (LBM) software
\citep{Michalski2021}.

In the previous ``state of the practice'' project, we measured 30 software
projects for each domain, but the measures were relatively shallow.  With this
re-boot we still target about 30 software examples from each domain, but we are
now collecting more data.  In keeping with the previous project, we still have
the constraint that the work load for applying the methodology to a given domain
needs to be feasible for a team as small as one individual, and for a time that
is short, ideally around a person month per domain.\footnote{A person month is
considered to be $20$ working days ($4$ weeks in a month, with $5$ days of work
per week) at $8$ person hours per day, or $20 \cdot 8 = 160$ person hours.}

To begin our re-boot of the previous methodology we critically assessed, and
subsequently modified, our
\href{https://github.com/adamlazz/DomainX/blob/master/TemplateToGradeSCSoft.pdf}
{previous set of measures}.  In addition, the following data has been added to
the new methodology:

\begin{itemize}
\item Characterization of the functionality provided by the software in the
  domain via a commonality analysis.
\item Project repository related data, such as the number of files, number of
  lines of code, percentage of issues that are closed, etc. 
\item Interviews with software developers in domain X.
\end{itemize}

Unlike the previous measurement process, the new methodology involves and
engages a domain expert partner throughout.  We did not previously engage the
domain expert with the rationale that we wished to eliminate potential bias.
However, this advantage is not worth the inability to evaluate the functionality
of the software.  Moreover, not having an expert makes navigating the on-line
resources difficult, since on-line resources are often silent on information
that is implicit to domain experts.  Furthermore, not every statement found
on-line will necessarily be accurate.  The importance of the domain expert is
particularly noteworthy when it comes time for publication and dissemination of
the state of the practice assessment.  Throughout this document the person (or
persons) that provide domain expertise will be designated as the \textit{Domain
Expert}.

In the proposed methodology, the collected data is combined to rank the software
within each domain using the Analytic Hierarchy Process (AHP) \citep{Saaty1980}.
As in the previous measurement exercise, we use AHP to develop a list of
software ranked by quality.  However, in the new process we do not stop with
this list.  The Domain Expert is consulted to verify the ordering, and to
discuss the decisions that led to the ranking.  The AHP process is used to
facilitate a conversation with the Domain Expert as a means to deepen our
understanding of the software in the domain, and the needs of typical
developers.

So that the collected data for a given domain can benefit the scientific
community, our recommendation is that all collected data be made public.  For
instance, the data collection for each domain can be put on a GitHub repository.
In addition to the project record left on GitHub, the final data can be exported
to \href{https://data.mendeley.com/}{Mendeley Data}.  As an example, the
measurements for the \href{https://data.mendeley.com/datasets/6kprpvv7r7/1}
{state of the practice for GIS software} are available on Mendeley.  Ideally the
full analysis of the state of the practice for domain X will also be published
in a suitable journal, allowing for dissemination/feedback and communication.

The scope of this methodology includes observations on product, artifact
(documentation, test scripts, etc.) and process quality for research software.
We leave the assessment of the performance of research software, for instance
using benchmarks, to other projects, such as the work of
\citet{kaagstrom1998gemm}.  Currently we are also leaving experiments to measure
usability and modifiability as future work, as discussed in
Section~\ref{SecFutureWork}.

Within this document the following typographic conventions have been adopted:
\begin{inparaenum}[i)] 
  \item \textcolor{blue}{Blue} text denotes a link to sections of this document,
  including citations 
  \item \textcolor{cyan}{Cyan} text denotes an external URL link.
\end{inparaenum}

The full methodology is presented in the sections that follow.
Section~\ref{SecResearchQuestions} highlights the research questions that are to
be answered for each measured domain. These questions are answered by the data
collected using the process outlined in Section~\ref{StepsAQDS}.  The major
steps in the process are outlined in Sections~\ref{SecIdentifyDomain} --
\ref{SecAHP}.  Following this, the time required for assessing a single domain
is estimated in Section~\ref{SecEstTimeRequired}.

\section{Research Questions} \label{SecResearchQuestions}

The following are the research questions that we wish to answer for each of our
selected domains.  In addition to answering the questions for each domain, we
also wish to combine the data from multiple domains to answer these questions
for research software in general.

\begin{enumerate}
\item What artifacts are present in current software packages? 
\item What tools (development, dependencies, project management) are used by
current software packages?
\item What principles, processes, and methodologies are used in the development
  of current software packages?
\item What are the pain points for developers working on research software
  projects?  What aspects of the existing processes, methodologies and tools do
  they consider as potentially needing improvement?  How should processes,
  methodologies and tools be changed to improve software development and
  software quality?
\item For research software developers, what specific actions are taken to
  address the following:
\begin{enumerate}
\item usability
\item traceability
\item modifiability
\item maintainability
\item correctness
\item understandability
\item unambiguity
\item reproducibility
\item visibility/transparency
\end{enumerate} 
\item How does software designated as high quality by this methodology compare
  with top rated software by the community?
\end{enumerate}

\section{Overview of Steps in Assessing Quality of the Domain Software}
\label{StepsAQDS}

To answer the above research questions (Section~\ref{SecResearchQuestions}), we
systematically measure the quality of the software through data collection.  An
overview of the measurement process is given in the following steps, starting
from determining a domain that is suitable for measurement:

\begin{enumerate}
\item Identify the domain (X). (Section~\ref{SecIdentifyDomain})
\item Ask the Domain Expert to create a top ten list of software packages in the
  domain. (Section~\ref{SecDomainExpert})
\item Meet with the Domain Expert to brief them on the overall objective,
  research proposal, research questions, measurement template, and developer
  interviews. (Section~\ref{SecDomainExpert}) \label{StepMeeting}
\item Identify the broad list of candidate software packages in the domain.
  (Section~\ref{SecIdentifyCandSoft})
\item Preliminary filter of software packages list.
  (Section~\ref{SecInitialFilter})
\item Review software list with Domain Expert. (Section~\ref{SecDomainExpert})
\item Domain Analysis (with the help of the Domain Expert).
(Section~\ref{SecDomainAnalysis})
\item Ask Domain Expert to vet domain analysis.
  (Section~\ref{SecDomainExpert})
\item Gather source code and documentation for each prospective software
  package.
\item Collect repository based information. (Section~\ref{SecEmpiricalMeasures})
\item Measure using measurement template. (Section~\ref{SecMeasureTemplate})
\item Survey developers. (Section~\ref{SecSurvey})
\item Use AHP process to rank the software packages. (Section~\ref{SecAHP})
\item Ask Domain Expert to vet AHP ranking. (Section~\ref{SecDomainExpert})
\item Answer research questions (from Section~\ref{SecResearchQuestions}) and
document answers.
\end{enumerate}


\section{How to Identify the Domain?} \label{SecIdentifyDomain} 

A domain of research software must be identified to begin the assessment.
Research software is defined in this exercise as ``software that is used to
generate, process or analyze results that [are intended] to appear in a
publication'' \citep{hettrick2014uk} in a scientific or engineering context.
Research software is a more general term for what is often called scientific
computing.  To be applicable for the methodology described in this document, the
chosen domain must have the following properties:

\begin{enumerate}
\item The domain must have well-defined and stable theoretical underpinning.  A
  good sign of this is the existence of standard textbooks, preferably
  understandable by an upper year undergraduate student.
\item There must be a community of people studying the domain.
\item The software packages must have open source options.
\item A preliminary search, or discussion with experts, suggests that there will
  be numerous, close to 30, candidate software packages in the domain that
  qualify as `research software'.
\end{enumerate}	

Some examples of domains that fit these criteria are finite element analysis
\citep{szabo1996finite}, quantum chemistry \citep{veryazov20042molcas},
seismology \citep{SmithEtAl2018}, as well as mesh generators
\citep{smith2016state}.

\section{Interaction With Domain Expert} \label{SecDomainExpert} 

As mentioned in the introduction (Section~\ref{SecIntroduction}), the Domain
Expert is an important member of the state of the practice assessment team.
Pitfalls exist if non-domain experts attempt to acquire an authoritative list of
software, or perform a commonality analysis, or try to definitively rank the
software. The main source of problems for non-domain experts is that they can
only rely on information that is available on-line, but on-line data has two
potential problems:
\begin{inparaenum}[i)]
  \item the on-line resources could have false or inaccurate information; and,
  \item the on-line resources could leave out relevant information that is so
in-grained with experts that nobody thinks to explicitly record the information.
\end{inparaenum}

Domain experts may be recruited from academia or industry.  The only
requirements are knowledge of the domain and a willingness to be engaged in the
assessment process.  The Domain Expert does not have to be a software developer,
but they should be a user of domain software.  Given that the domain experts are
likely to be busy people, the measurement process cannot put to much of a burden
on their time.

In advance of the first meeting with the Domain Expert (Step~\ref{StepsAQDS} in
Section~\ref{StepsAQDS}) the expert is asked to create a top ten list of
software packages in the domain.  This is done to help the expert get in the
right mind set in advance of the first meeting.  Moreover, by doing the exercise
in advance, we avoid the potential pitfall of the expert approving the
discovered list of software without giving it adequate thought.  The emphasis
during the first meeting is for the Domain Expert to learn what is expected of
them.  The discussion should also cover avenues for publication and
dissemination.

The Domain Experts are asked to vet the collected data and analysis.  In
particular, they are asked to vet the proposed list of software packages, the
domain analysis and the AHP ranking.  These interactions can be done either
electronically or with in-person (or virtual) meetings.  

\section{How to Identify Candidate Software?} \label{SecIdentifyCandSoft}

Once the domain of interest is identified, the candidate software for measuring
can be found through search engine queries targeting authoritative lists of
software.  Potential places to search include \href{https://github.com/}
{GitHub}, \href{https://swmath.org/} {swMATH} and domain related publications,
such as review articles. Domain Experts are also asked for their suggestions and
are asked to review the initial draft of the software list.  

When forming the list and reviewing the candidate software the following
properties should be considered:

\begin{enumerate}
\item The software functionality must fall within the identified domain.
\item The source code must be viewable.
\item The empirical measures listed in Section \ref{SecEmpiricalMeasures} should
  ideally be available, which implies a preference for GitHub-style
  repositories.
\item The software cannot be marked as incomplete, or in an initial development
  phase.
\end{enumerate}

\section{How to Initially Filter the Software List?} \label{SecInitialFilter}

If the list of software is too long (over around 30 packages), then steps need
to be taken to create a more manageable list. To reduce the length of the list,
the following filters are applied.  The filters are applied in the priority
order listed, with the filtering process stopped once the list size is
manageable.

\begin{enumerate}
\item Scope: Software is removed by narrowing what functionality is considered
  to be within the scope of the domain.
\item Usage: Software packages are eliminated if their installation procedure is
  not clear and easy to follow.
\item Age: The older software packages (age being measured by the last date when
  a change was made) are eliminated, except in the cases where an older software
  package appears to be highly recommended and currently in use.  (The Domain
  Expert can be consulted on this question, if necessary.)
\end{enumerate}

Copies of both the initial and filtered lists, along with the rationale for
shortening the list, should be kept for traceability purposes.

\section{Domain Analysis} \label{SecDomainAnalysis}

Since each domain we will study will have a reasonably small scope, we will be
able to view the software as constituting a program family.  The concept of a
program family is defined by \citet{parnas1976design} as ``a set of programs
whose common properties are so extensive that it is advantageous to study the
common properties of the programs before analyzing individual members''.
Studying the common properties within a family of related programs is termed a
domain analysis.

The domain analysis consists of a commonality analysis of the family of software
packages. Its purpose is to show the relationships between these packages, and
to facilitate an understanding of the informal specification and development of
them. \cite{weiss1997defining} defines commonality analysis as an approach to
defining a family by identifying commonalities, variabilities, and common
terminology for the family. Commonalities are goals, theories, models,
definitions and assumptions that are common between family members.
Variabilities are goals, theories, models, definitions and assumptions that
differ between family members. Associated with each variability are its
parameters of variation, which summarize the possible values for that
variability, along with their potential binding time.  The binding time is when
the value of the variability is set.  It could be set as specification time,
build time (when the program is being compiled) or run time (when the code is
executing).

The final result of the domain analysis will be tables of commonalities,
variabilities, and parameters of variation of a program family.
\cite{smith2008commonality} present a template for conducting a commonality
analysis, which was referred to when conducting this work.
\cite{weiss1998commonality} describes another commonality analysis technique for
deciding the members of a program family. \cite{SmithAndChen2004} and
\cite{SmithMcCutchanAndCarette2017} are examples of a commonality analysis for a
family of mesh generating software and a family of material models,
respectively. The steps to produce a commonality analysis are:

\begin{enumerate}
\item Write an Introduction
\item Write an Overview of the Domain Knowledge
\item List Commonalities
\item List Variabilities
\item List Parameters of Variation
\item Add Terminology, Definitions, Acronyms
\end{enumerate}


\section{Repository Based Measures} \label{SecEmpiricalMeasures}

Some quality measurements rely on gathering raw and processed data from software
repositories. We focus on data that is reasonably easy to collect, which we
combine and analyze. The measures that are collected relate to the research
questions (Section~\ref{SecResearchQuestions}). For instance, we collect data to
see how large a project is, to ascertain a project’s popularity, and to
determine whether to project is being actively developed.

Section~\ref{rawdata} lists the raw data that is collected.  Some of this data
can be observed from GitHub repository metrics. The rest can be collected using
freeware tools. \href{https://github.com/tomgi/git_stats}{GitStats} is used to
measure the number of binary files as well as the number of added and deleted
lines in a repository. This tool is also used to measure the number of commits
over different intervals of time. \href{https://github.com/boyter/scc}{Sloc Cloc
and Code (scc)} is used to measure the number of text based files as well as the
number of total, code, comment, and blank lines in a repository. These tools
were selected due to their installability, usability, and ability to gather the
empirical measures listed below. Details on installing and running the tools can
be found in Appendix~\ref{SecRepoTools}.  Section \ref{processeddata} introduces
the required processed data, which is calculated using the raw data.

\subsection{Raw Data}\label{rawdata}

The following raw data measures are extracted from repositories:

\begin{itemize}
\item Number of stars.
\item Number of forks.
\item Number of people watching the repository.
\item Number of open pull requests.
\item Number of closed pull requests.	
\item Number of developers.	
\item Number of open issues.
\item Number of closed issues.
\item Initial release date.
\item Last commit date.
\item Programming languages used.
\item Number of text-based files.
\item Number of total lines in text-based files.
\item Number of code lines in text-based files.
\item Number of comment lines in text-based files.
\item Number of blank lines in text-based files.
\item Number of binary files.  
\item Number of total lines added to text-based files.
\item Number of total lines deleted from text-based files.
\item Number of total commits.
\item Numbers of commits by year in the last 5 years. (Count from as early as
  possible if the project is younger than 5 years.)
\item Numbers of commits by month in the last 12 months.
\end{itemize}

\subsection{Processed Data}\label{processeddata}

The following measures are calculated from the raw data:

\begin{itemize}
\item Status of software package as either dead or alive, where alive is defined
  as the presence of repository commits or software package version releases in
  the last 18 months.
\item Percentage of identified issues that are closed.
\item Percentage of code that is comments.
\end{itemize}

\noindent The time frame of 18 months was selected as the separating point
between alive and dead projects because this is the usual timeframe for
operating system updates.

\section{Measure Using Measurement Template} \label{SecMeasureTemplate}

\begin{figure}[!ht]
	\begin{center}
	  \includegraphics[width=1.0\textwidth]{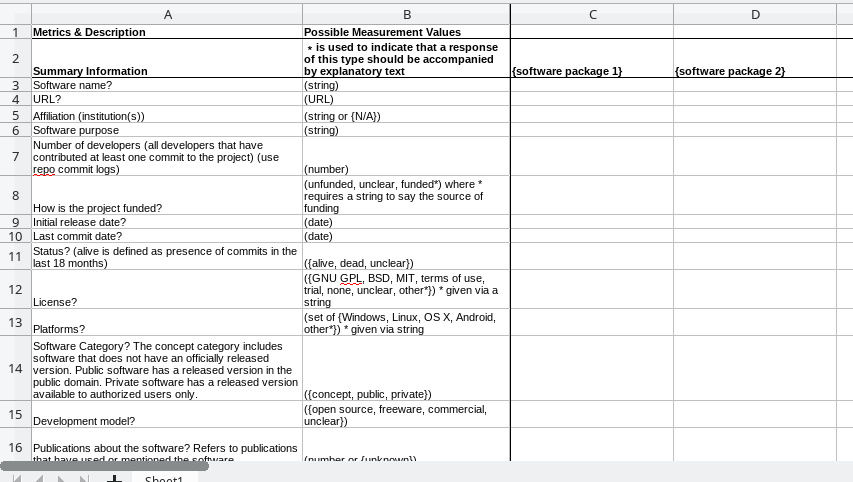}
	  \caption{Excerpt of the Top Section of the Measurement Template (Summary Information)}
	  \label{measurement_template_image}
	\end{center}
  \end{figure}

The Measurement Template is found in Appendix~\ref{SecGradingTemplate}.  This
template is used to track measurements and quality scores for all of the
software packages in the domain. For each software package, we fill-in the
template questions. This process can take between 1 to 4 hours for each package.
Project developers can be contacted for help regarding installation, if
necessary, but a cap of about 2 hours should be imposed on the entire
installation process, to keep the overall measurement time feasible.  To save
time, a
\href{https://github.com/smiths/AIMSS/blob/master/StateOfPractice/Methodology/Combined_MeasurementTemplate_EmpiricalMeasures.xlsx}{blank
measurement template spreadsheet} has been prepared, with the measures as rows.
An excerpt of the spreadsheet is shown in
Figure~\ref{measurement_template_image}.  A column should be added to this
template for each software package to be measured.

The full template consists of 108 questions categorized under 9 qualities.  The
questions were designed to be unambiguous, quantifiable and measurable with
limited time and domain knowledge. The measures are grouped under headings for
each quality, and one for summary information. The summary information (shown in
Figure~\ref{measurement_template_image}) is the first section of the template.
This section summarizes general information, such as the software name, number
of developers, etc.  We follow the definitions given by
\citet{GewaltigAndCannon2012} for the software categories.  Public means software
intended for public use.  Private means software aimed only at a specific
group, while the concept category is used for software written simply to demonstrate
algorithms or concepts. The three categories of development models are: open
source, where source code is freely available under an open source license;
free-ware, where a binary or executable is provided for free; and, commercial,
where the user must pay for the software product.  

Following the summary section are sections to measure 9 qualities: 
\begin{inparaenum}
	\item installability;
	\item correctness and verifiability;
	\item surface reliability;
	\item surface robustness;
	\item surface usability;
	\item maintainability;
	\item reusability;
	\item surface understandability; and,
	\item visibility/transparency. 
\end{inparaenum} 
Definitions of these qualities are available in a
\href{https://github.com/smiths/AIMSS/blob/master/StateOfPractice/QDefOfQualities/QDefOfQualities.pdf}{working
document on software quality}.  Several of the qualities use the word
``surface''.  This is to highlight that, for these qualities in particular, the
best that we can do is a shallow measure of the quality.  For instance, we are
not currently doing any experiments to measure usability.  Instead, we are
looking for an indication that usability was considered by the developers.  We
do this by looking for cues in the documentation, like a getting started manual,
a user manual and documentation of expected user characteristics.

Most of the data to be collected should be straightforward from reviewing the
measurement template.  However, in a few cases extra guidance is necessary
to eliminate ambiguity, as follows:

\begin{enumerate}
\item Initial release date: Mark the release year if an exact date is not
  available.
\item Publications about the software: A list of publications can be found
  directly on the website of some software packages. For others use Google
  Scholar or a similar index.
\item Is there evidence that performance was considered?: Search the software
  artifacts for any mention of speed, storage, throughput, performance
  optimization, parallelism, multi-core processing, or similar
  considerations. The search function on GitHub can help.
\item Getting started tutorial: Sometimes this is found within another artifact,
  like the user manual.
\item Continuous integration: Search the software artifacts for any mention of
  continuous integration. The search function on GitHub can help.  In some
  cases, \texttt{yaml} files will provide a hint that continuous integration is
  employed.
\end{enumerate}

\noindent To fill-in the spreadsheet template, the following steps should be followed:

\begin{enumerate}
\item Gather the summary information into the top section of the document
  (Figure \ref{measurement_template_image}).
\item Using the GitStats tool that is described in
  Section~\ref{SecEmpiricalMeasures} gather the measurements for the Repo
  Metrics - GitStats section found near the bottom of the spreadsheet.
\item Using the SCC tool that is also described in
  Section~\ref{SecEmpiricalMeasures} gather the measurements for the Repo
  Metrics -- SCC section found near the bottom of the spreadsheet.
\item If the software package is found on git, gather the measurements for the
  Repo Metrics - the GitHub section found near the bottom of the spreadsheet.
\item Review installation documentation and attempt to install the software
  package on a virtual machine.
\item Gather the measurements for installability
\item Gather the measurements for correctness and verifiability
\item Gather the measurements for surface reliability
\item Gather the measurements for surface robustness
\item Gather the measurements for surface usability
\item Gather the measurements for maintainability
\item Gather the measurements for reusability
\item Gather the measurements for surface understandability
\item Gather the measurements for visibility and transparency
\item Assign a score out of ten for each quality. The score can be measured
  using the Measurement Template Impression Calculator, found in
  Appendix~\ref{SecImpressionCalculator}. For each quality measurement, the file
  indicates the appropriate score to assign the measurement based on possible
  measurement values.
\end{enumerate}

As in \citet{SmithEtAl2016}, Virtual machines (VMs) are used to provide an
optimal testing environments for each package. VMs were used because it is
easier to start with a fresh environment without having to worry about existing
libraries and conflicts. Moreover, when the tests are complete the VM can be
deleted, without any impact on the host operating system. The most significant
advantage of using VMs is to level the playing field. Every software install
starts from a clean slate, which removes ``works-on-my-computer'' errors. When
filling in the measurement template spreadsheet, the the details for each VM
should be noted, including hypervisor and operating system version.

\section{Survey Developers} \label{SecSurvey}

In the previous state of the practice measurement process
\citep{SmithEtAl2018_arXivGIS, SmithEtAl2016, SmithEtAl2018}, we only based our
assessment on information available in on-line software repos.  However, this
approach meant we weren't able to learn about the development process, the
attitudes of the developers, the pain points for developers and how the
developers handle modifiability, reproducibility and usability.  Therefore, in
the reboot of the measurement process, we have explicitly added a stage for
interviewing research software developers.

We designed a list of 20 questions to guide our interviews, which can be found
in Appendix~\ref{SecSurveyQuestions}. Some questions are about the background of
the software, the development teams, the interviewees, and how they organize the
projects. We also ask about the developer's understandings of the users. Some
questions focus on the current and past difficulties, and the solutions the team
has found, or will try. We also discuss the importance and current situations of
documentation. A few questions are about specific software qualities, such as
maintainability, understandability, usability, and reproducibility. The
interviews are semi-structured based on the question list; we ask follow-up
questions when necessary. Based on our experience, the interviewees usually
bring up some exciting ideas that we did not expect, and it is worth expanding
on these topics.

Our methodology suggests requesting interviews with a developer from each of the
30 software package.  Requests for interviews are sent to all packages so as to
not cause a potential bias by singling out any subset of the list. Moreover,
since not every developer will agree to the interview request, asking 30 times
will typically yield a reasonable number of responses. In our experience, the
response rate is between 15\% and 30\%.  In some cases multiple developers from
the same project will agree to be interviewed. When sending out interview
requests, we recommend finding the contacts on the projects’ website, or code
repository, or publications, or the biographic pages of the teams’ institutions.
We send at most two interview request emails to a contact for each software
package.  Meeting will typically be held using on-line meeting software, like
Zoom or Teams.  This facilitates recording and automatic transcription of the
meetings.

The interviewees should follow a process where they can make informed consent.
The interviews should follow standard ethics guideline of asking for consent
before interviewing, recording, and including participant details in the report.
The interview process presented here was approved by the McMaster University
Research Ethics Board under the application number 
\href{https://github.com/smiths/AIMSS/blob/master/StateOfPractice/MACREM/Application.pdf}
{MREB\#: 5219}.

\section{Analytic Hierarchy Process} \label{SecAHP}

The Analytical Hierarchy Process (AHP) is a decision-making technique that can
be used when comparing multiple options by multiple criteria. AHP focuses on
pair-wise comparisons between all options for all criteria.  The advantage of
pair-wise comparisons is that they facilitates a separation of concerns.  Rather
than worry about the entire problem, the decision maker can focus on one
comparison at a time.  In our work AHP is used for comparing and ranking the
software packages of a domain using the quality scores that are gathered in the
Measurement Template (Appendix~\ref{SecGradingTemplate}). AHP performs a
pairwise analysis between each of the 9 quality options for each of the 30
software packages.  This results in a matrix, which is used to generate an
overall score for each software package for the given criteria.
\cite{SmithEtAl2016} shows how AHP is applied to ranking software based on
quality measures. We have developed a
\href{https://github.com/smiths/AIMSS/blob/master/StateOfPractice/AHP2020/LBM/README.txt}{tool}
for conducting this process. The tool includes an AHP JAR script and a
sensitivity analysis JAR script that is used to ensure that the software package
rankings are appropriate with respect to the uncertainty of the quality scores.
The
\href{https://github.com/smiths/AIMSS/blob/master/StateOfPractice/AHP2020/LBM/README.txt}{README
file} outlines the requirements for, and configuration and usage of, the JAR
scripts. The JAR scripts, source code, and required libraries are located in the
same folder as the README file.



\section{Estimate of Time Required} \label{SecEstTimeRequired}

Table~\ref{TabPersonHours} estimates the time required (in person hours) to
complete a state of the practice assessment for domain X.  The table assumes
that the domain has already been decided and the Domain Expert has been
recruited.  The time spent by the Domain Expert is not included in the numbers
shown in the table, since the amount of time that the domain expert will work
independently of the rest of the assessment team will be small.  Moreover, this
amount of time will vary greatly depending on the preferred work habits of the
Domain Expert.  The table follows the steps outlined in Section~\ref{StepsAQDS}.
Time is not included for reviewing the methodology. Moreover, it is assumed that
the template spreadsheets linked in this document, and the developed AHP tool,
will be employed, rather than developing new tools.  The person hours given are
a rough estimate, based on our experience completing assessments for medical
image analysis software \citep{Dong2021} and for Lattice Boltzmann Method (LBM)
software \citep{Michalski2021}.  These two domains were assessed at the same
time as designing the methodology presented in this document.  We did our best
to estimate the time spent on measurement and separate it from the time spend on
design and development.  The estimate assumes 30 software packages will be
measured; the numbers will need to be adjusted if the total packages changes.

\begin{table}[h]
  \caption{Estimated Person Hours for Assessing the State of Practice for Domain
  X} \label{TabPersonHours}
  \centering
  \begin{tabular}{p{10cm} l}
    \toprule
    \textbf{Task} & \textbf{Hours} \\
    \midrule

    Initial 1 hour meeting with the Domain Expert plus meeting prep & 5
    \addtocounter{totHours}{5} \\

    Identify broad list of candidate software
    (Section~\ref{SecIdentifyCandSoft}) & 12 \addtocounter{totHours}{12} \\

    Filter software list (Section~\ref{SecInitialFilter}) (10 minutes per
    package for 30 packages) & 5 \addtocounter{totHours}{5} \\

	Review software list with Domain Expert (Section~\ref{SecDomainExpert}) & 2
	\addtocounter{totHours}{2} \\

    Domain analysis (with help of Domain Expert)
    (Section~\ref{SecDomainAnalysis}) & 20 \addtocounter{totHours}{20} \\

	Vet domain analysis with Domain Expert (Section~\ref{SecDomainExpert}) & 3
	\addtocounter{totHours}{3} \\

	Gather source code and documentation for each package (10 minutes per
	package for 30 packages)) & 5 \addtocounter{totHours}{5} \\

    Collect repository based data (Section~\ref{SecEmpiricalMeasures}) (10
    minutes per package for 30 packages) & 5 \addtocounter{totHours}{5} \\

	Measure using measurement template (Section~\ref{SecMeasureTemplate}) (2.5
    hours per repo for 30 repos) & 75 \addtocounter{totHours}{75} \\

    Solicit developers for interviews & 2 \addtocounter{totHours}{2} \\

    Conduct interviews (1.5 hour interviews with 10 developers (assuming 1 in 3
    developers agree to an interview)) & 15 \addtocounter{totHours}{15} \\

    AHP ranking & 2 \addtocounter{totHours}{2} \\

	Work with Domain Expert to vet AHP ranking & 2 \addtocounter{totHours}{2} \\

    Analyze data and answer research questions & 20 \addtocounter{totHours}{20}\\

    \midrule
    \textbf{Total} & \textbf{\thetotHours} \\
    \bottomrule
  \end{tabular}
  
\end{table}  

The total number of person hours is \total{totHours} hours.  This is close to
our goal of 1 month of person hours (160 hours).  The amount of time spent by
the Domain Expert can be estimated by summing the Domain Expert items in
Table~\ref{TabPersonHours} and adding an estimate of the time that they will
independently spend on their assigned tasks.  If we assume that the Domain
Expert will spend 2 hours on the domain analysis and another 2 hours with
answering questions, the Domain Expert time will be about 12 person hours.

\section{Future Work} \label{SecFutureWork}

As explained in the introduction (Section~\ref{SecIntroduction}), our eventual
goal is to improve the state of practice of software development, which requires
a baseline means for measuring the current state of the practice.  Now that we
have a methodology for assessing the state of practice for a given domain, the
next task is to complete the measurements for multiple domains.  Using the
previous and updated methodologies, we have measured the state of
practice for the following domains: Geographic Information Systems
\citep{SmithEtAl2018_arXivGIS}, Mesh Generators \citep{SmithEtAl2016},
Seismology software \citep{SmithEtAl2018}, Statistical software for psychology
\citep{SmithEtAl2018_StatSoft}, medical image analysis software \citep{Dong2021}
and for Lattice Boltzmann Method software \citep{Michalski2021}.  Future domains
for measurement could include finite element software, computational medicine,
machine learning, ordinary differential equation solvers, computer graphics,
stoichiometry, etc.

With the wealth of data from assessing the state of practice for multiple
domains, the next step is a meta-analysis.  We would look at how the different
domains compare. What lessons from one domain could be applied in other domains?
What (if any) differences exist in the pain points between domains?  Are there
differences in the tools, processes, and documentation between domains?

The current methodology is constrained by limited resources.  A 4 hour cap on
the measurement time for each software package limits what can be assessed.
Within this limit, we can't measure some important qualities, like usability and
modifiability.  In the future, we propose a more time-consuming process that
would capture these other quality measures.  To improve the feasibility, the
more time consuming measurements would not have to be completed for all 30
packages. Instead, a short list could be identified using the output of the AHP
ranking to select the top projects, or to select a sample of interesting
projects across the quality spectrum.

If we can add measures for modifiability and usability, we can start to measure
the quality impact of software development processes, tools and techniques.  For
instance, we have a project (called
\href{https://github.com/JacquesCarette/Drasil} {Drasil}
\citep{SzymczakEtAl2016}), which facilitates a development process that focuses
on knowledge capture, followed by generation of code and other software
artifacts from the captured knowledge.  To understand the advantages and
disadvantages of Drasil, we could measure its quality.  In particular, we would
like to see the impact of a generative process on the quality of usability and
modifiability.

\subsection{Measuring Usability in the Future}

In the future, we propose an experiment for assessing the usability of a given
software package.  Some initial thoughts on how this might be done are recorded
in this section.  To do the experiment we need an experimental subject, who will
be required to complete tasks with the software being studied.  The interaction
with the software will allow the study subject to experience the software's
usability.  The tasks for the subject to complete would vary by domain;
therefore, the tasks would be selected with the help of the Domain Expert.
Criteria for selecting candidate tasks are as follows:

\begin {enumerate}
\item Tasks should be executable for subjects with novice to intermediate
experience.
\item All tasks should take no more than one hour.
\item Tasks should include the basic/common use cases of the software package.
\item Include tasks that require sequential or hierarchical steps for
completion.
\end {enumerate}

Once the study subject has experience with the software, they will be in a
position to judge its usability.  We will measure the usability using a
standardized usability questionnaire, like the
\href{https://www.usabilitest.com/sus-pdf-generator} {System Usability Scale
questionnaire} or the
\href{https://uiuxtrend.com/pssuq-post-study-system-usability-questionnaire/}
{Post-Study System Usability Questionnaire}.

As a starting point for the experiment design, the procedure could be something
like the following:

\begin {enumerate}
\item Survey participants to collect pre-experiment data (background, experience
of subjects (especially with domain software)).
\item Participants perform tasks based on task defined by Domain Experts.
\item Observe the study subjects (take notes, record sessions (screen
recorder), watch for body languages and verbal cues).
\item Survey the study subjects to collect feedback (post-experiment interview),
complete usability questionnaire.
\item Prepare a summary report of the experimental results.
\end {enumerate}

\subsection{Measuring Modifiability in the Future}

The next experiment is designed to gather qualitative data regarding the
modifiability of each software package. This proposed experiment also involves
experimental subjects/participants, who in this case are asked to modify a set
of software packages.  The specific modifications requested will again depend on
the software domain.  In advance of the experiment the Domain Expert will be
asked the likely changes for software in this domain.  We emphasize likely
changes, instead of any changes because software cannot be designed so that
everything is equally easy to change \citep{parnas1986rational}.  The procedure
could be along the following lines:

\begin{enumerate}
	\item Domain Expert lists all likely changes that a developer
	might be asked to make in a software package in the domain.
	\item Survey study participants to collect pre-experiment data (background,
	experience of subjects (especially with domain software)).
	\item Participants perform modification tasks for likely changes on each
	software package being studied.
	\item Observe the study subjects (take notes, record sessions (screen
	recorder), watch for body languages and verbal cues).
	\item Record time needed to make the changes.	
	\item Confirm through testing that the modified software has the correct
	behaviour.
	\item Survey the study subjects to collect feedback (post-experiment interview).
	\item Prepare a summary report of the experimental results.
\end{enumerate}

The study subjects should make the same changes in multiple pieces of software.
The reporting of the results will focus more on the relative time differences
between the set of software packages, rather than the absolute time to make any
given change. To remove biases caused by the participants experience, the
different study subjects should use a different order as they go through the
list of software packages.  Details for this experiment still need to be
resolved, such as how to take the participants prior knowledge into account,
especially with respect to their programming experience.

\section{Concluding Remarks} \label{SecConcludingRemarks}

We have outlined a methodology for assessing the state of the practice for any
given research software domain.  (Although the scope of the current work has been
on research software, there is little in the methodology that is specific to
research software, except for the interview question related to the quality of
reproducibility.)  When applying the methodology to a given domain, we provide a
means to answer the following questions:
\begin{inparaenum}[i)]
\item What artifacts (documents, code, test cases, etc.) are present?
\item What tools are used?
\item What principles, process and methodologies are used?
\item What are the pain points for developers?
\item What actions are used to improve qualities like maintainability and
reproducibility?
\item What specific actions are taken to achieve the qualities of usability,
traceability, modifiability, maintainability, correctness, understandability,
unambiguity, reproducibility and visibility/transparency?
\item How does software designated as high quality by this methodology compare
  with top rated software by the community?
\end{inparaenum} 

The methodology depends on the engagement of a Domain Expert.  The Domain
Expert's role is to ensure that the assessment is consistent with the culture of
the community of practitioners in the domain.  The Domain Expert also has an
important role to play with the domain analysis.  For each domain we 
conduct a domain analysis to look at the commonalities, variabilities and
parameters of variation, for the family of software in the domain.  The domain
analysis means that software can be compared not just based on its quality, but
also based on its functionality.

The methodology follows a systematic procedure that begins with identifying the
domain and ends with answering the research questions posed above.  In between
we collect an authoritative list of about 30 software packages.  For each
package in the list we fill in our measurement template.  The template consists
of repository related data (like number of open issues, number of lines of code,
etc.) and 108 measures/questions related to 9 qualities: installability,
correctness/verifiability, reliability, robustness, usability, maintainability,
reusability, understandability and visibility/transparency. Filling in the
template requires installing the software, running simple tests (like completing
the getting started instructions (if present)), and searching the code,
documentation and test files.

The data for each domain is used to rank the software package according to each
quality dimension using AHP.  The ranking is not intended to identify a single
best software package.  Instead the ranking is intended to provide insights on
the top set of software for each quality.  The top performers can be contrasted
with the lesser performers to gain insight into what practices in the domain are
working.  Deeper insight can be obtained by combining this data with the
interview data from asking each recruited developer 20 questions.

Combining the quantitative data from the measurement template with the interview
results, along with the domain experts knowledge, we can determine the current
state of the practice for domain X.  Using our methodology, spreadsheet
templates and AHP tool, we estimate (based on our experience with using the
process) the time to complete an assessment for a given domain at
\total{totHours} person hours.

\newpage

\appendix

\section{A Guide to Repository-Based Measurement Tools} \label{SecRepoTools}

This appendix covers the tools used for collecting repository based data
(Section~\ref{SecEmpiricalMeasures}).  The two tools covered are git\_stats and
scc.  The tools do not have to be used in any particular order.

\subsection{git\_stats}

\subsubsection{Introduction}

Source Code: \href{https://github.com/tomgi/git_stats}{GitHub repo}

\subsubsection{User Manual} \label{git_stats_manual}

Official Manual: \href{https://github.com/tomgi/git_stats}{GitHub repo}

\subsubsection{Demo of Installation and Running the Tool}

The installation steps on your machine may be different from this section.
Please refer to the user manual mentioned in Section \ref{git_stats_manual}, if
necessary.  The steps shown here were executed on a virtual machine with 8 cores
and 16 GB RAM running Debian GNU/Linux 9.11.

\begin{enumerate}
\item Install ruby/gem environment
\begin{lstlisting}
apt-get install ruby ruby-nokogiri ruby-nokogiri-diff ruby-nokogumbo
\end{lstlisting}

Check the installation:
\begin{lstlisting}
gem --version
\end{lstlisting}
    
\item Install the tool
\begin{lstlisting}
sudo gem install git_stats
\end{lstlisting}
    
\item Prepare the target repo

Make sure the target repo (the repo to be analyzed, not the repo of this tool)
is on your machine.
In this demo, the target repo is downloaded from a GitHub repo:
\begin{lstlisting}
# change [git path] to the url of your target repo
git clone [git path]
# e.g. git clone https://github.com/nroduit/Weasis.git
\end{lstlisting}
    
\item Generate analytics
\begin{lstlisting}
# make sure [repo path] is the target repo path
# the [output path] can be anywhere you desire
git_stats generate -p [repo path] -o [output path]
# e.g. git_stats generate -p /home/user/git-stats/Weasis -o 
# /home/user/git-stats/Weasis-analytics
\end{lstlisting}
    
\item View the analytics

View the analytic results by open [output path]/index.html with any browser or
other software supporting HTML web page format.

\begin{figure}[!ht]
\includegraphics[scale=0.4]{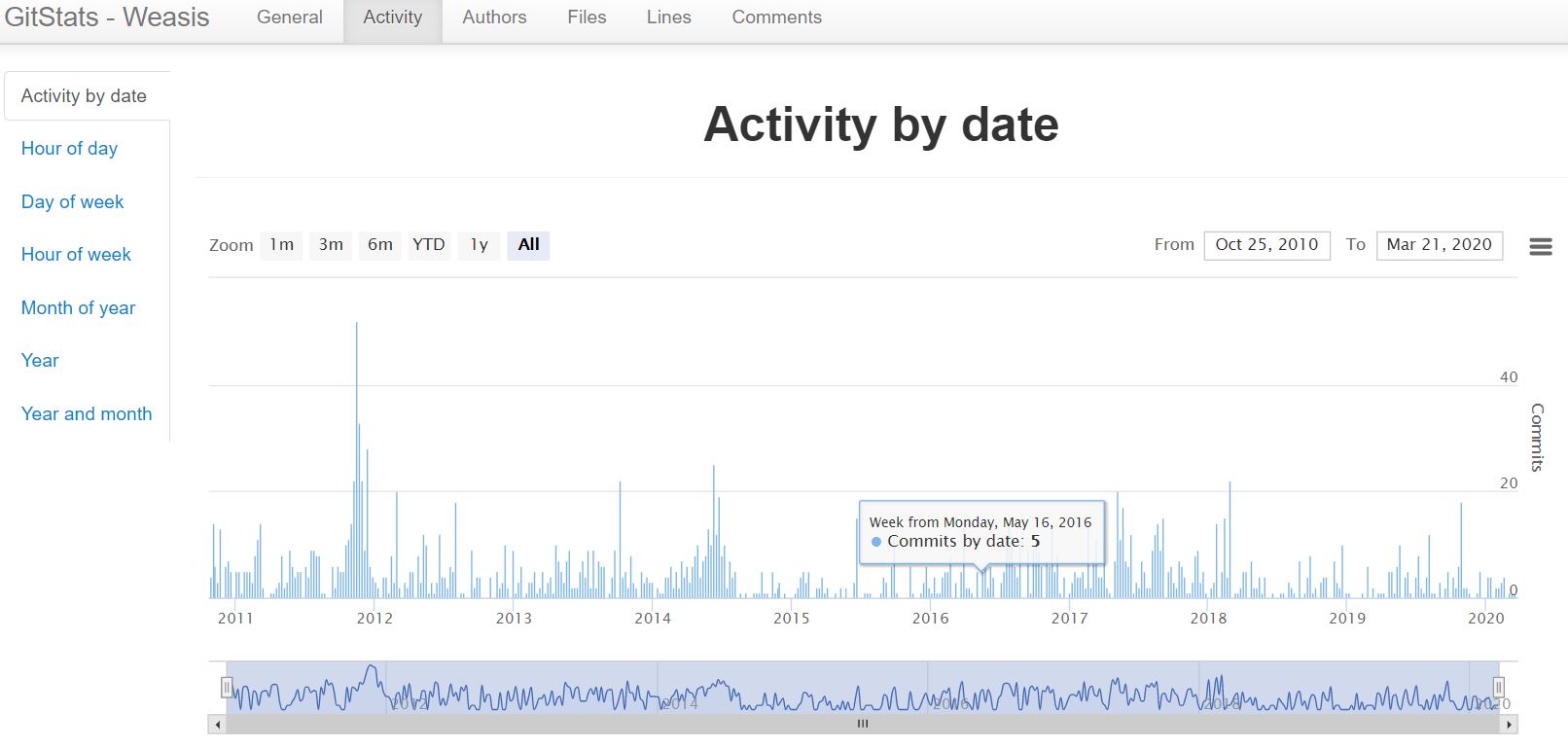}
\end{figure}

\item Download the data

On most of the taps of this web page, the data can be downloaded for more
analytics by clicking the menu button beside the data-range section.

\end{enumerate}

\subsection{scc}

\subsubsection{Introduction}

Source Code: \href{https://github.com/boyter/scc}{GitHub repo}

\subsubsection{User Manual} \label{scc_manual}

Official Manual: \href{https://github.com/boyter/scc}{GitHub repo}

\subsubsection{Demo of Installation and Running the Tool}

The installation steps on your machine may be different from this section.
Please refer to the user manual mentioned in Section \ref{scc_manual}, if
necessary.  These steps were executed on a virtual machine with 8 cores and 16
GB RAM running Debian GNU/Linux 9.11

\begin{enumerate}
\item Install Golang

Follow the \href{https://golang.org/doc/install}{official instructions}, or the
following demo,

download the installation package:
\begin{lstlisting}
wget https://dl.google.com/go/go1.14.3.linux-amd64.tar.gz
\end{lstlisting}
    
unpack to /usr/local:
\begin{lstlisting}
sudo tar -C /usr/local -xzf go1.14.3.linux-amd64.tar.gz
\end{lstlisting}

use a text editor to open \textasciitilde/.profile, e.g.:
\begin{lstlisting}
nano ~/.profile
\end{lstlisting}

add the following lines to the end of this file:
\begin{lstlisting}
export GOPATH=$HOME/go
export PATH=$PATH:/usr/local/go/bin:$GOPATH/bin
\end{lstlisting}

save the file, and load the commands into the current shell instance:
\begin{lstlisting}
source ~/.profile
\end{lstlisting}

check the installation:
\begin{lstlisting}
go version
\end{lstlisting}

\item Install the tool
\begin{lstlisting}
go get -u github.com/boyter/scc/
\end{lstlisting}
    
\item Prepare the target repo

Make sure the target repo (the repo to be analyzed, not the repo of this tool)
is on your machine.
In this demo, the target repo is downloaded from a GitHub repo:
\begin{lstlisting}
# change [project path] to your desired folder
cd [project path]
git clone https://github.com/nroduit/Weasis.git
\end{lstlisting}

\item Generate analytics
\begin{lstlisting}
# make sure [repo path] is the target repo path
cd [repo path]
# use scc to generate analytics
scc
\end{lstlisting}
    
\item View the analytics

The results will be appear something like the following.

\begin{figure}[!ht]
\includegraphics[scale=0.5]{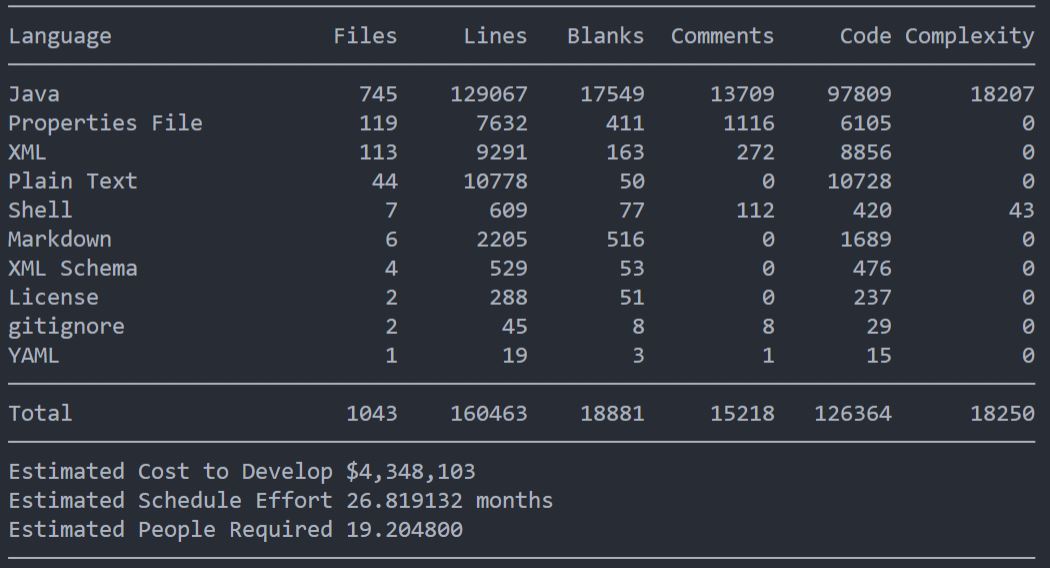}
\end{figure}

\end{enumerate}

\newpage

\section{Measurement Template} \label{SecGradingTemplate}

The table below lists the full set of measures that are assessed for each
software product.  The measures are grouped under headings for each quality, and
one for summary information.  Following each measure, the type for a valid
result is given in brackets.  Many of the types are given as enumerated sets.
For instance, the response on many of the questions is one of ``yes,'' ``no,''
or ``unclear.''  The type ``number'' means natural number, a positive integer.  The
types for date and url are not explicitly defined, but they are what one would
expect from their names.  In some cases the response for a given question is not
necessarily limited to one answer, such as the question on what platforms are
supported by the software product.  Case like this are indicated by ``set of''
preceding the type of an individual answer.  The type in these cases are then
the power set of the individual response type.  In some cases a superscript $^*$
is used to indicate that a response of this type should be accompanied by
explanatory text.  For instance, if problems were caused by uninstall, the
reviewer should note what problems were caused.  
The template also include 3 sections at the end for summarizing the repository
based metrics.  A
\href{https://github.com/smiths/AIMSS/blob/master/StateOfPractice/Methodology/Combined_MeasurementTemplate_EmpiricalMeasures.xlsx}
{blank measurement template spreadsheet} is available to save time with data
entry.

\newpage

\captionof{table}{Measurement Template}\label{measurementtemplate}
\def\arraystretch{1.22}
\begin{tabular}{p{14cm}}
	\hline
	\textbf{Summary Information}\\
	\hline
	Software name? (string)\\
	URL? (URL)\\
	Affiliation (institution(s)) (string or {N/A})\\
	Software purpose (string)\\
	Number of developers (all developers that have contributed at least one commit
	to the project) (use repo commit logs) (number)\\
	How is the project funded? (unfunded, unclear, funded$^*$) where $^*$ requires a
	string to say the source of funding\\
	Initial release date? (date)\\
	Last commit date? (date)\\
	Status? (alive is defined as presence of commits in the last 18 months)
	({alive, dead, unclear})\\
	License? ({GNU GPL, BSD, MIT, terms of use, trial, none, unclear, other$^*$}) $^*$
	given via a string \\
	Platforms? (set of {Windows, Linux, OS X, Android, other$^*$}) $^*$ given via
	string\\
	Software Category? The concept category includes software that does not have
	an officially released version. Public software has a released version in the
	public domain. Private software has a released version available to authorized
	users only. ({concept, public, private})\\
	Development model? ({open source, freeware, commercial, unclear})\\
	Publications about the software? Refers to publications that have used or
	mentioned the software. (number or {unknown})\\
	Source code URL? ({set of url, n/a, unclear})\\
	Programming language(s)? (set of {FORTRAN, Matlab, C, C++, Java, R, Ruby,
	Python, Cython, BASIC, Pascal, IDL, unclear, other$^*$}) $^*$ given via string \\
	Is there evidence that performance was considered? Performance refers to
	either speed, storage, or throughput. ({yes$^*$, no})\\
	Additional comments? (can cover any metrics you feel are missing, or any other
	thoughts you have) \\
	\hline
\end{tabular}

\def\arraystretch{1.5}
\begin{tabular}{p{14cm}}
	\hline		
	\textbf{Installability (Measured via installation on a virtual machine.) }\\
	\hline
	Are there installation instructions? ({yes, no})\\
	Are the installation instructions in one place? Place referring to a single
	document or web-page. ({yes, no, n/a})\\
	Are the installation instructions linear? Linear meaning progressing  in a
	single series of steps. ({yes, no, n/a})\\
	Are the instructions written as if the person doing the installation has none
	of the dependent packages installed? ({yes, no, unclear})\\
	Are compatible operating system versions listed? ({yes, no})\\
	Is there something in place to automate the installation (makefile, script,
	installer, etc)? ({yes$^*$, no})\\
	If the software installation broke, was a descriptive error message displayed?
	({yes, no, n/a})\\
	Is there a specified way to validate the installation? ({yes$^*$, no})\\
	How many steps were involved in the installation? (Includes manual steps like
	unzipping files) Specify OS. (number, OS)\\
	What OS was used for the installation? ({Windows, Linux, OS X, Android, other$^*$
	}) $^*$given via string\\
	How many extra software packages need to be installed before or during
	installation? (number)\\
	Are required package versions listed? ({yes, no, n/a})\\
	Are there instructions for the installation of required packages /
	dependencies? ({yes, no, n/a})\\
	Run uninstall, if available. Were any obvious problems caused? ({yes$^*$ , no,
	unavail})\\
	Overall impression? ({1 .. 10})\\
	Additional comments? (can cover any metrics you feel are missing, or any other
	thoughts you have)\\
	\hline
\end{tabular}

\def\arraystretch{1.33}
\begin{tabular}{p{14cm}}
	\hline		
	\textbf{Correctness and Verifiability}\\
	\hline
	Any reference to the requirements specifications of the program or theory
	manuals? ({yes$^*$ , no, unclear})\\
	What tools or techniques are used to build confidence of correctness?
	({literate programming, automated testing, symbolic execution, model checking,
	assertions used in the code, Sphinx, Doxygen, Javadoc, confluence, unclear,
	other$^*$}) $^*$ given via string\\
	If there is a getting started tutorial? ({yes, no})\\
	Are the tutorial instructions linear? ({yes, no, n/a})\\
	Does the getting started tutorial provide an expected output? ({yes, no$^*$,
	n/a})\\
	Does your tutorial output match the expected output? ({yes, no, n/a})\\
	Are unit tests available?  ({yes, no, unclear})\\
	Is there evidence of continuous integration? (for example mentioned in
	documentation, Jenkins, Travis CI, Bamboo, other) ({yes$^*$, no, unclear})\\
	Overall impression? ({1 .. 10})\\
	Additional comments? (can cover any metrics you feel are missing, or any other
	thoughts you have) \\
	\hline	
	\textbf{Surface Reliability}\\
	\hline
	Did the software “break” during installation? ({yes$^*$ , no})\\
	If the software installation broke, was the installation instance recoverable?
	({yes, no, n/a})\\
	Did the software “break” during the initial tutorial testing? ({yes$^*$, no,
	n/a})\\
	If the tutorial testing broke, was a descriptive error message displayed?
	({yes, no, n/a})\\
	If the tutorial testing broke, was the tutorial testing instance recoverable?
	({yes, no, n/a})\\
	Overall impression? ({1 .. 10})\\
	Additional comments? (can cover any metrics you feel are missing, or any other
	thoughts you have)\\
	\hline		
\end{tabular}

\def\arraystretch{1.4}
\begin{tabular}{p{14cm}}
	\hline		
	\textbf{Surface Robustness}\\
	\hline
	Does the software handle unexpected/unanticipated input (like data of the
	wrong type, empty input, missing files or links) reasonably? (a reasonable
	response can include an appropriate error message.) ({yes, no$^*$ })\\
	For any plain text input files, if all new lines are replaced with new lines
	and carriage returns, will the software handle this gracefully? ({yes, no$^*$,
	n/a})\\
	Overall impression? ({1 .. 10})\\
	Additional comments? (can cover any metrics you feel are missing, or any other
	thoughts you have)\\
	\hline		
	\textbf{Surface Usability}\\
	\hline
	Is there a getting started tutorial? ({yes, no})\\
	Is there a user manual? ({yes, no})\\
	Are expected user characteristics documented? ({yes, no})\\
	What is the user support model? FAQ? User forum? E-mail address to direct
	questions? Etc. (string)\\
	Overall impression? ({1 .. 10})\\
	Additional comments? (can cover any metrics you feel are missing, or any other
	thoughts you have)\\
	\hline
\end{tabular}

\def\arraystretch{1.4}
\begin{tabular}{p{14cm}}
	\hline	
	\textbf{Maintainability}\\
	\hline
	What is the current version number? (number)\\
	Is there any information on how code is reviewed, or how to contribute?
	({yes$^*$, no})\\
	Are artifacts available? (List every type of file that is not a code file –
	for examples please look at the ‘Artifact Name’ column of
	https://gitlab.cas.mcmaster.ca/SEforSC/se4sc/-/blob/git-svn/GradStudents/Olu/ResearchProposal/Artifacts\_MiningV3.xlsx)
	({yes$^*$, no, unclear}) $^*$list via string\\
	What issue tracking tool is employed? (set of {Trac, JIRA, Redmine, e-mail,
	discussion board, sourceforge, google code, git, BitBucket, none, unclear,
	other$^*$}) $^*$ given via string\\
	What is the percentage of identified issues that are closed? (percentage)\\
	What percentage of code is comments? (percentage)\\
	Which version control system is in use? ({svn, cvs, git, Github, unclear,
	other$^*$}) $^*$ given via string\\
	Overall impression? ({1 .. 10})\\
	Additional comments? (can cover any metrics you feel are missing, or any other
	thoughts you have)\\
	\hline		
	\textbf{Reusability}\\
	\hline
	How many code files are there? (number)\\
	Is API documented? ({yes, no, n/a})\\
	Overall impression? ({1 .. 10})\\
	Additional comments? (can cover any metrics you feel are missing, or any other
	thoughts you have)\\
	\hline		
\end{tabular}

\def\arraystretch{1.4}
\begin{tabular}{p{14cm}}
	\hline		
	\textbf{Surface Understandability (Based on 10 random source files)}\\
	\hline
	Consistent indentation and formatting style? ({yes, no, n/a})\\
	Explicit identification of a coding standard? ({yes$^*$, no, n/a})\\
	Are the code identifiers consistent, distinctive, and meaningful? ({yes, no$^*$ ,
	n/a})\\
	Are constants (other than 0 and 1) hard coded into the program? ({yes, no$^*$ ,
	n/a})\\
	Comments are clear, indicate what is being done, not how? ({yes, no$^*$ , n/a})\\
	Parameters are in the same order for all functions? ({yes, no$^*$ , n/a})\\
	Is the name/URL of any algorithms used mentioned? ({yes, no$^*$ , n/a})\\
	Is code modularized? ({yes, no$^*$ , n/a})\\
	Overall impression? ({1 .. 10})\\
	Additional comments? (can cover any metrics you feel are missing, or any other
	thoughts you have)\\
	\hline		
	\textbf{Visibility/Transparency}\\
	\hline
	Is the development process defined? If yes, what process is used. ({yes$^*$, no,
	n/a})\\
	Are there any documents recording the development process and status?  ({yes$^*$,
	no}))\\
	Is the development environment documented? ({yes$^*$, no})\\
	Are there release notes? ({yes$^*$, no})\\
	Overall impression? ({1 .. 10})\\
	Additional comments? (can cover any metrics you feel are missing, or any other
	thoughts you have)\\
	\hline		
	\textbf{Raw Metrics (Measured via git\_stats)}\\
	\hline
	Number of text-based files. (number)\\
	Number of binary files. (number)\\
	Number of total lines in text-based files. (number)\\
	Number of total lines added to text-based files. (number)\\
	Number of total lines deleted from text-based files. (number)\\
	Number of total commits. (number)\\
	Numbers of commits by year in the last 5 years. (Count from as early as
	possible if the project is younger than 5 years.) (list of numbers)\\
	Numbers of commits by month in the last 12 months. (list of numbers)\\
	\hline
\end{tabular}

\def\arraystretch{1.4}
\begin{tabular}{p{14cm}}
\hline		
\textbf{Raw Metrics (Measured via scc)}\\
\hline
Number of text-based files. (number)\\
Number of total lines in text-based files. (number)\\
Number of code lines in text-based files. (number)\\
Number of comment lines in text-based files. (number)\\
Number of blank lines in text-based files. (number)\\
\hline
\textbf{Repo Metrics (Measured via GitHub)}\\
\hline
Number of stars. (number)\\
Number of forks. (number)\\
Number of people watching this repo. (number)\\
Number of open pull requests. (number)\\
Number of closed pull requests. (number)\\
\hline
\end{tabular}

\newpage

\section{Measurement Template Impression Calculator} \label{SecImpressionCalculator}

The table below lists how each quality measure of the measurement template is
used to calculate an overall impression in each software quality set.

\newpage

\captionof{table}{Measurement Template}
\def\arraystretch{1.5}
\begin{tabular}{p{14cm}}
	\hline		
	\textbf{Installability (Measured via installation on a virtual machine.) }\\
	\hline
	Are there installation instructions? ({yes=1, no=-1})\\
	Are the installation instructions in one place? Place referring to a single
	document or web-page. ({yes=1, no=0, n/a=0})\\
	Are the installation instructions linear? Linear meaning progressing  in a
	single series of steps. ({yes=1, no=0, n/a=0})\\
	Are the instructions written as if the person doing the installation has none
	of the dependent packages installed? ({yes=1, no=0, unclear=0})\\
	Are compatible operating system versions listed? ({yes=1, no=0})\\
	Is there something in place to automate the installation (makefile, script,
	installer, etc)? ({yes$^*$=1, no=-1})\\
	If the software installation broke, was a descriptive error message displayed?
	({yes=0, no=-2, n/a=1})\\
	Is there a specified way to validate the installation? ({yes$^*$=1, no=0})\\
	How many steps were involved in the installation? (Includes manual steps like
	unzipping files) Specify OS. ($<$10 = 1)\\
	What OS was used for the installation? (does not count)\\
	How many extra software packages need to be installed before or during
	installation? ($<$10 = 1)\\
	Are required package versions listed? ({yes=1, no=0, n/a=1})\\
	Are there instructions for the installation of required packages /
	dependencies? ({yes=1, no=0, n/a=1})\\
	Run uninstall, if available. Were any obvious problems caused? ({yes$^*$=0, no=1,
	unavail=1})\\
	Overall impression? (a sum of $>$10 is rounded down to 10)\\
	\hline
\end{tabular}

\def\arraystretch{1.33}
\begin{tabular}{p{14cm}}
	\hline		
	\textbf{Correctness and Verifiability}\\
	\hline
	Any reference to the requirements specifications of the program or theory
	manuals? ({yes$^*$=2, no=0, unclear=0})\\
	What tools or techniques are used to build confidence of correctness? (any=1,
	unclear=0)\\
	If there is a getting started tutorial? ({yes=2, no=0})\\
	Are the tutorial instructions linear? ({yes=1, no=0, n/a=0})\\
	Does the getting started tutorial provide an expected output? ({yes=1, no$^*$=0,
	n/a=0})\\
	Does your tutorial output match the expected output? ({yes=1, no=0, n/a=0})\\
	Are unit tests available?  ({yes=1, no=0, unclear=0})\\
	Is there evidence of continuous integration? (for example mentioned in
	documentation, Jenkins, Travis CI, Bamboo, other) ({yes$^*$=1, no=0,
	unclear=0})\\
	\hline	
	\textbf{Surface Reliability}\\
	\hline
	Did the software “break” during installation? ({yes$^*$=0, no=5})\\
	If the software installation broke, was the installation instance recoverable?
	({yes=5, no=0, n/a=0})\\
	Did the software “break” during the initial tutorial testing? ({yes$^*$=0, no=5,
	n/a=0})\\
	If the tutorial testing broke, was a descriptive error message displayed?
	({yes=2, no=0, n/a=0})\\
	If the tutorial testing broke, was the tutorial testing instance recoverable?
	({yes=3, no=0, n/a=0})\\
	\hline		
	\textbf{Surface Robustness}\\
	\hline
	Does the software handle unexpected/unanticipated input (like data of the
	wrong type, empty input, missing files or links) reasonably? (a reasonable
	response can include an appropriate error message.) ({yes=5, no$^*$=0})\\
	For any plain text input files, if all new lines are replaced with new lines
	and carriage returns, will the software handle this gracefully? ({yes=5,
	no$^*$=0, n/a=5})\\
	\hline		
\end{tabular}

\def\arraystretch{1.4}
\begin{tabular}{p{14cm}}
		\hline		
	\textbf{Surface Usability}\\
	\hline
	Is there a getting started tutorial? ({yes=3, no=0})\\
	Is there a user manual? ({yes=4, no=0})\\
	Are expected user characteristics documented? ({yes=1, no=0})\\
	What is the user support model? FAQ? User forum? E-mail address to direct
	questions? Etc. (one=1, two+=2, none=0)\\
	\hline
	\textbf{Maintainability}\\
	\hline
	What is the current version number? (provided=1, nothing=0)\\
	Is there any information on how code is reviewed, or how to contribute?
	({yes$^*$=1, no=0})\\
	Are artifacts available? (List every type of file that is not a code file –
	for examples please look at the ‘Artifact Name’ column of
	https://gitlab.cas.mcmaster.ca/SEforSC/se4sc/-/blob/git-svn/GradStudents/Olu/ResearchProposal/Artifacts\_MiningV3.xlsx)
	(Rate 0 – 2 depending on how many and perceived quality)\\
	What issue tracking tool is employed? (nothing=0, email of other private=1,
	anything public or accessible by all devs (eg git) = 2)\\
	What is the percentage of identified issues that are closed? (50$\%$+=1,
	$<$50$\%$=0)\\
	What percentage of code is comments? (10$\%$+=1, $<$10$\%$=0)\\
	Which version control system is in use? (anything=2, nothing=0)\\
	\hline		
	\textbf{Reusability}\\
	\hline
	How many code files are there? (0-9=0, 10-49=1, 50-99=3, 100-299=4, 300-599=5,
	600-999=6, 1000+=8)\\
	Is API documented? ({yes=2, no=0, n/a=0})\\
	\hline
\end{tabular}

\def\arraystretch{1.4}
\begin{tabular}{p{14cm}}
\hline	
\textbf{Surface Understandability (Based on 10 random source files)}\\
\hline
Consistent indentation and formatting style? ({yes=1, no=0, n/a=0})\\
Explicit identification of a coding standard? ({yes$^*$=1, no=0, n/a=0})\\
Are the code identifiers consistent, distinctive, and meaningful? ({yes=2,
no$^*$=0, n/a=0})\\
Are constants (other than 0 and 1) hard coded into the program? ({yes=1, no$^*$=0,
n/a=0})\\
Comments are clear, indicate what is being done, not how? ({yes=2, no$^*$=0,
n/a=0})\\
Is the name/URL of any algorithms used mentioned? ({yes=1, no$^*$=0, n/a=0})\\
Parameters are in the same order for all functions? ({yes=1, no$^*$=0, n/a=0})\\
Is code modularized? ({yes=1, no$^*$=0, n/a=0})\\
\hline		
\textbf{Visibility/Transparency}\\
\hline
Is the development process defined? If yes, what process is used. ({yes$^*$=3,
no=0, n/a=0})\\
Are there any documents recording the development process and status?  ({yes$^*$=3,
no=0}))\\
Is the development environment documented? ({yes$^*$=2, no=0})\\
Are there release notes? ({yes$^*$=2, no=0})\\
\hline
\end{tabular}

\newpage

\section{Survey Questions for Developers} \label{SecSurveyQuestions}

Interviews will be one-to-one and will be open-ended (not just “yes or no”
answers). Because of this, the exact wording of the questions may change a
little. If more information, or clarification, is needed during the
conversation, follow up questions will be asked, such as: ``So, you are saying
that ...?'', ``Please tell me more?'', or ``Why do you think that is?''

\subsection{Information about the developers and users} 

These questions are based on questions used in \citet{Jegatheesan2016}

\begin{enumerate}
\item Interviewees' current position/title? degrees?
\item Interviewees' contribution to/relationship with the software?
\item Length of time the interviewee has been involved with this software?
\item How large is the development group?
\item Do you have a defined process for accepting new contributions into your team?
\item What is the typical background of a developer?
\item What is your estimated number of users? How did you come up with that estimate?
\item What is the typical background of a user?
\end{enumerate}

\subsection{Information about the software}

Square brackets are used for traceability to the  relevant research
questions found in Section~\ref{SecResearchQuestions}.

\begin{enumerate}
\item Currently, what are the most significant obstacles in your development
process?
\item How might you change your development process to remove or reduce these
obstacles?
\item How does documentation fit into your development process? Would improved
documentation help with the obstacles you typically face? [research question 5b (traceability),
research question 5i (visibility/transparency)]
\item In the past, is there any major obstacle to your development process that
has been solved? How did you solve it?
\item What is your software development model? For example, waterfall, agile,
etc.
\item What is your project management process? Do you think improve this process
can tackle the current problem? Were any project management tools used?
\item Was it hard to ensure the correctness of the software? If there were any
 obstacles, what methods have been considered or practiced to improve the
 situation? If practiced, did it work? [research question 5e (correctness)]
\item When designing the software, did you consider the ease of future changes?
 For example, will it be hard to change the structure of the system, modules or
 code blocks? What measures have been taken to ensure the ease of future changes
 and maintains? [research question 5d (maintainability), 
 research question 5c (modifiability)]
\item Provide instances where users have misunderstood the software. What, if
any, actions were taken to address understandability issues? [research question
5f (understandability)]
\item What, if any, actions were taken to address usability issues? [
research question 5a (usability)]
\item Do you think the current documentation can clearly convey all necessary
knowledge to the users? If yes, how did you successfully achieve it? If no, what
improvements are needed? [research question 5g (unambiguity)]
\item Do you have any concern that your computational results won't be
reproducible in the future? Have you taken any steps to ensure reproducibility?
[research question 5h (reproducibility)]
\end{enumerate}

\newpage

\bibliographystyle {plainnat}
\bibliography {Methodology_arXiv_version}

\end{document}